\documentclass[3p,times,twocolumn]{elsarticle}

\usepackage{ecrc}
\usepackage{multicol}

\volume{00}

\firstpage{1}

\journalname{Nuclear Physics B Proceedings Supplement}

\runauth{}

\jid{nuphbp}

\jnltitlelogo{Nuclear Physics B Proceedings Supplement}

\usepackage{amssymb}
\usepackage{amsmath}

\newcommand{\be}{\begin{equation}}
\newcommand{\ee}{\end{equation}}
\newcommand{\bea}{\begin{eqnarray}}
\newcommand{\eea}{\end{eqnarray}}

\newcommand{\pup}{p^\uparrow}

\newcommand{\ZP}[1]{{\it Z.\ Phys.}\ {\bf #1}}

\newcommand{\PR}[1]{{\it Phys.\ Rev.}\ {\bf #1}}

\begin{document}
\title{\bf Transverse Single Spin Asymmetries and Charmonium Production}
\begin{frontmatter}



\author{Rohini M. Godbole\fnref{label1}}
\ead{rohini@cts.iisc.ernet.in}
\author{Anuradha Misra\fnref{label2}\corref{cor2}}
\ead{anuradha.misra@gmail.com}
\author{Asmita Mukherjee\fnref{label3}}
\ead{asmita@phy.iitb.ac.in}
\author{Vaibhav S. Rawoot\fnref{label2}}
\ead{vaibhavrawoot@gmail.com}

\address[label1]{Centre for High Energy Physics, Indian Institute of Science, Bangalore, India-560012}
\address[label2]{Department of Physics,University of Mumbai, \\Santa Cruz(E), Mumbai, India-400098}
\address[label3]{Department of Physics,Indian Institute of Technology, \\Bombay, Mumbai, India-400076}
 \cortext[cor2]{Speaker of the talk}

\dochead{}





\begin{abstract}
We estimate transverse spin single spin asymmetry(TSSA) in the process $e+p^\uparrow \rightarrow J/\psi +X$ using 
color evaporation model of charmonium production. We take into account transverse momentum dependent(TMD) 
evolution of Sivers function and parton distribution function and show that the there is a reduction in the asymmetry 
as compared to our earlier estimates wherein the $Q^2$ - evolution was implemented only through DGLAP evolution 
of unpolarized gluon densities. 
\end{abstract}

\begin{keyword}
Single Spin Asymmetry, Charmonium

\end{keyword}

\end{frontmatter}


\section{Introduction}

Single spin asymmetries(SSA's)   arise  in the scattering of transversely polarized nucleons 
off an unpolarized nucleon (or virtual photon) target, when  the 
final state  hadrons  have asymmetric distribution in the transverse plane
perpendicular to the beam direction.   SSA for inclusive process $A^\uparrow+B \rightarrow C+X$
depends  on the polarization vector of the scattering hadron A  and is defined by
\begin{equation} 
A_N = \frac{d\sigma ^\uparrow \, - \, d\sigma ^\downarrow}
{d\sigma ^\uparrow \, + \, d\sigma ^\downarrow} \label{an} 
\end{equation}
 
 Non - zero SSAs have been observed  over the years- in pion production at Fermilab\cite{AdamsBravar1991}
 and at RHIC\cite{KruegerAllogower1999} in $p\pup$ collisions as well as in Semi-inclusive deep inelastic scattering (SIDIS) 
 experiments at HERMES\cite{Hermes} and COMPASS\cite{Compass}. These results have generated a lot of 
 interest amongst theoreticians to investigate the mechanism involved and to  understand the 
 underlying physics. 

The initial attempts to provide theoretical predictions  of asymmetry, based on collinear factorization 
of pQCD,  led to estimates which were 
too small as compared to the experimental results\cite{alesio-review}.  
 In  collinear factorization formalism, the parton distribution functions (PDF's) and fragmentation functions (FF's) are  integrated over intrinsic transverse momentum of the partons and hence depend only on longitudinal momentum fraction x.  The observation that SSAs calculated within collinear formalism were almost vanishing suggested that  these asymmetries may be due to  
parton's transverse motion and spin orbit correlation. A generalization of 
factorization theorem, in the form of transverse momentum dependent (TMD) factorization which includes the  transverse momentum dependence of 
PDF's and FF's,  was proposed as a possible approach to account for  the asymmetries\cite{Sivers1990}.  

 One of the TMD PDF's of  interest is Sivers function, which gives  the probability of finding an unpolarized quark inside a transversely polarized 
 proton. 
The Sivers function, $\Delta^N f_{a/p\uparrow}(x,{\bf k_{\perp a}})$,  defined by 
\begin{eqnarray}
\Delta^N f_{a/p\uparrow}(x,{\bf k_{\perp a}}) \equiv
\hat f_{a/p\uparrow}(x, {\bf k}_{\perp a})-\hat f_{a/\downarrow}(x, {\bf k}_{\perp a})\nonumber\\
=  \hat f_{a/\uparrow}(x, {\bf k}_{\perp a})-\hat f_{a/\uparrow}(x, - {\bf k}_{\perp a})  
\label{delf1} 
\end{eqnarray} 
 is related to the number density of partons inside a proton with transverse polarization
S, three momentum ${\bf p}$ and intrinsic transverse momentum ${\bf k}_\perp$ of partons,
and its spin dependence is given  by 
\begin{equation}
 \Delta^N f_{a/p\uparrow}(x,{\bf k_{\perp a}})= \Delta^N f_{a/p\uparrow}(x, k_{\perp}) \> 
{\bf S} \cdot ({\bf \hat p} \times {\bf  \hat k}_{\perp}) 
\label{delf2}
\end{equation}

Parametrizations of quark Sivers distributions have been obtained from fits of SSA in SIDIS experiments\cite{Anselmino:2008sga}.  However, not much information is available on  
 gluon Sivers function. Processes that have been studied with the aim of getting information about this TMD are back to back correlations in azimuthal angles of of jet produced in $p p^\uparrow$ scattering\cite{Boer-PRD69(2004)094025}
and D meson production in $p p^\uparrow$ scattering\cite{Anselmino2004}. Heavy quark and quarkonium systems have also been proposed as  natural probes to study gluon Sivers function due to the fact that  the production is sensitive 
to intrinsic transverse momentum especially at low momentum\cite{feng}. It has been suggested, in the context of  deep inelastic 
scattering\cite{Brodsky2002}  that the initial and final state interactions may lead to non-vanishing SSAs. Single transverse spin asymmetry in heavy quarkonium production in lepton-nucleon
and nucleon-nucleon collisions has been investigated by Yuan {\it etal}  taking into account the initial and final state interactions
\cite{feng} 
 and it has been shown  that the asymmetry is very sensitive to the production mechanism. The three main models of heavy quarkonium production,  which have been proposed and tested in unpolarized scattering,  are Color Singlet Model\cite{sing}, Color Evaporation Model (CEM)\cite{hal, cem0} and the NRQCD factorization approach\cite{octet}. It was argued in Ref.\cite{feng} that  the asymmetry  should be non-zero in ep collisions only in color-octet model and in pp collisions only in color-singlet model. Thus, SSA in charmonium production can be used to throw some light on  the issue of production mechanism.
In this work, we present estimates of SSA in the process $e+p^\uparrow\rightarrow J/\psi +X$ and compare the results obtained using TMD evolution of PDF's with our earlier results which were obtained using DGLAP evolution only. 

\section{Transverse Single Spin Asymmetry in $e+p^\uparrow\rightarrow J/\psi +X$}
The  first estimate of SSA in photoproduction (i.e. low virtuality electroproduction) of $J/\psi$  in the scattering of 
electrons off transversely polarized protons were provided by us in Ref.\cite{Godbole:2012bx} using Color Evaporation Model.
In the process under consideration, at LO, there is contribution only from a single
partonic subprocess and therefore, it can be used as a clean probe of gluon Sivers function.

Color Evaporation Model (CEM) was introduced in 1977 by Fritsch and was revived in 1996 by Halzen\cite{hal}. 
This model gives a good description of photoproduction data after inclusion of higher order QCD corrections\cite{cem1} 
and also of the hadroproduction CDF data \cite{cem2} after inclusion of  $k_T$ smearing.
In CEM,  
the cross-section for a quarkonium state H is some fraction $F_H$ of the cross-section 
for producing $Q\bar{Q}$ pair with invariant mass below the $M\bar{M}$ threshold,
where M is the lowest mass meson containing the heavy quark Q:
\begin{align}
\sigma_{CEM}[h_A h_B\rightarrow H+X]=F_H\sum_{i,j}\int_{4m^2}^{4m_M^2}  d\hat{s}  \nonumber \\
\times \int dx_1 dx_2\>f_i(x_1,\mu)\>f_j(x_2,\mu)\>\hat{\sigma}_{ij}(\hat{s})\> \delta(\hat{s}-x_1x_2s) 
\end{align}

We have used a  generalization of CEM expression for electroproduction of $J/\psi$ by 
taking into account the transverse momentum dependence
of the  gluon distribution function and the William Weizsacker (WW) function which gives the photon distribution of the electron in equivalent photon approximation\cite{wwf1}.  
The cross section for the process  $e+p^\uparrow\rightarrow J/\psi +X$ is  then given by 
\begin{align}
\sigma^{e+p^\uparrow\rightarrow e+J/\psi + X}=
\int_{4m_c^2}^{4m_D^2} dM_{c\bar c}^2  dx_\gamma dx_g [d^2{\bf k}_{\perp\gamma}d^2{\bf k}_{\perp g}]\nonumber\\
\times f_{g/p^{\uparrow}}(x_g,{\bf k}_{\perp g}) 
f_{\gamma/e}(x_{\gamma},{\bf k}_{\perp\gamma})\>
\frac{d\hat{\sigma}^{\gamma g\rightarrow c\bar{c}}}{dM_{c\bar c}^2} 
\label{dxec-ep}
\end{align}
where $ f_{\gamma/e}(x_{\gamma},{\bf k}_{\perp\gamma})$ is the distribution function of the photon in the electron. We assume  a gaussian 
form for the ${\bf k}_\perp$ dependence of pdf's \cite{Anselmino:2008sga}, 
\begin{equation}
f(x,k_{\bot})=f(x)\frac{1}{\pi\langle k^{2}_{\bot}\rangle} 
e^{-k^{2}_{\bot}/\langle{k^{2}_{\bot}\rangle}} 
\label{gauss2}
\end{equation}
where $\langle k^{2}_{\bot}\rangle=0.25 GeV^2$.  $ f_{\gamma/e}(x_{\gamma},{\bf k}_{\perp\gamma})$ is also assumed to have a similar ${\bf k}_\perp$ dependence and is given by 
\begin{equation}
f_{\gamma/e}(x_\gamma,k_{\perp \gamma})=f_{\gamma/e}(x_\gamma)\frac{1}{\pi\langle k^{2}_{\perp \gamma}\rangle} 
e^{-k^{2}_{\perp \gamma}/\langle{k^{2}_{\perp \gamma}\rangle}}. 
\label{gauss-g}
\end{equation} 
where $f_{\gamma/e}(x_\gamma)$ is the William Weizsacker function given by \cite{ww}:
\begin{eqnarray}
f_{\gamma /e}(y,E)=\frac{\alpha}{\pi} \frac{1+(1-y)^2}{y}\left(ln\frac{E}{m}-\frac{1}{2}\right)\nonumber \\
+\frac{y}{2}\left[ln\left(\frac{2}{y}-2\right)+1\right] 
+\frac{(2-y)^2}{2y}ln\left(\frac{2-2y}{2-y}\right) 
\label{ww-function}
\end{eqnarray}
y being the energy fraction of the electron carried by the photon. 

Using Eq.~\ref{delf2}, the expression for the numerator of the asymmetry reduces to \cite{Godbole:2012bx}
\begin{align}
\frac{d^3\sigma^\uparrow}{dy d^2{\bf q}_T}-\frac{d^3\sigma^\downarrow}{dy d^2{\bf q}_T}=
\frac{1}{2}\int_{4m^2_c}^{4m^2_D}[dM^2]\int [dx_{\gamma} dx_g]  \nonumber\\ 
\times \int [d^2{\bf k}_{\perp\gamma} d^2{\bf k}_{\perp g}]
\Delta^{N}f_{g/p^{\uparrow}}(x_{g},{\bf k}_{\perp g})    \nonumber\\
\times f_{\gamma/e}(x_{\gamma},{\bf k}_{\perp\gamma}) \,
\delta^4(p_g+p_{\gamma}-q)
{\hat \sigma}_0^{\gamma g \rightarrow c{\bar c}}(M^2) 
\label{nssa}\end{align}
where $q=p_c+p_{\bar c}$ 
and  $\hat{\sigma_{0}}^{\gamma g\rightarrow c\bar{c}}(M^2)$ is the partonic cross section\cite{gr78}:

\begin{align}
\hat{\sigma_{0}}^{\gamma g\rightarrow c\bar{c}}(M^2)=&
\frac{1}{2}e_{c}^2\frac{4\pi\alpha\alpha_s}{M^2} \bigg[(1+\gamma-\frac{1}{2}\gamma^2)
\ln{\frac{1+\sqrt{1-\gamma}}{1-\sqrt{1-\gamma}}}  \nonumber \\
&-(1+\gamma)\sqrt{1-\gamma}\bigg]
\end{align}
$\gamma=4m_c^2/M^2$
and  ${\bf q}_T$ and ${\bf k}_\perp$ are the transverse momenta of the gluon and $J/\psi$ respectively with 
azimuthal angles $\phi_q$ and $\phi_{k_\perp}$:
\begin{equation}
\bf q_T = q_T(\cos\phi_{q}, \, \sin\phi_{q},\, 0)  \nonumber
\end{equation}
\begin{equation}
\bf k_{\perp} = k_{\perp}(\cos\phi_{k_\perp}, \, \sin\phi_{k_\perp}, \, 0)
\end{equation}
The mixed product ${\bf S} \cdot (\hat{\bf p} \times \hat{\bf k}_{\perp})$ in $\Delta^{N}f_{g/p^{\uparrow}}(x_{g},{\bf k}_{\perp g})  $
gives an azimuthal dependence of the form, 
\begin{equation}
{\bf S}\cdot (\hat{\bf p}\times \hat{\bf k}_\perp)=\cos{\phi_{k_\perp}} 
\end{equation}
Taking $sin(\phi_{q}-\phi_S)$  as a weight\cite{vogelsang-weight}, the asymmetry integrated over the azimuthal angle of 
$J/\psi$ is given by 
\begin{equation}
A_N^{sin(\phi_{q}-\phi_S)} =\frac{\int d\phi_{q}
[d\sigma ^\uparrow \, - \, d\sigma ^\downarrow]sin(\phi_{q}-\phi_S)}
{\int d\phi_{q}[d\sigma ^\uparrow \, + \, d\sigma ^\downarrow]} 
\end{equation} 
which finally leads to 
\begin{eqnarray}\label{an2}
 A_N =\frac{\int d\phi_{q} [\int_{4m^2_c}^{4m^2_D}[dM^{2}]\int[d^2{\bf k}_{\perp g}]
\Delta^{N}f_{g/p\uparrow}(x_{g},{\bf k}_{\perp g})f_{\gamma/e}(x_{\gamma},{\bf q}_T-{\bf k}_{\perp g})\hat\sigma_{0}]sin(\phi_{q}-\phi_S)}
{2\int d\phi_{q}[\int_{4m^2_c}^{4m^2_D}[dM^{2}]\int[d^{2}{\bf k}_{\perp g}]f_{g/P}(x_g,{\bf k}_{\perp g})
f_{\gamma/e}(x_{\gamma},{\bf q}_T-{\bf k}_{\perp g})
{\hat \sigma}_0 ]} 
\end{eqnarray}
where
\begin{equation}
d\sigma=
\frac{d^3\sigma}{dy \, d^2\bf q_T}, \quad\quad x_{g,\gamma} = \frac{M}{\sqrt s} e^{\pm y}  \nonumber
\end{equation}

\section{Models for Sivers function}
In our analysis, we have used the following parameterization for the gluon Sivers function\cite{Anselmino:2008sga}
\begin{align}
 \Delta^Nf_{g/p\uparrow}(x,{\bf k}_\perp) = 2 {\cal N}_g(x)\,h({\bf k}_\perp)\,f_{g/p}(x) \nonumber \\
\times \frac{e^{-k^{2}_{\bot}/\langle{k^{2}_{\bot}\rangle}}}{\pi\langle k^{2}_{\bot}\rangle} 
\cos{\phi_{k_\perp}} 
\label{dnf}
\end{align}
where ${\mathcal N}_g(x)$ is an  x dependent normalization. 
We have used two different models for the  functional forms of  $h(k_\perp)$:  
In Model(1)\cite{Anselmino2009}\\
\begin{equation}
h(k_\perp) = \sqrt{2e}\,\frac{k_\perp}{M_{1}}\,e^{-{{\bf k}_\perp^2}/{M_{1}^2}}
\label{siverskt}
\end{equation}
whereas in Model(2)\cite{Anselmino2004}
\begin{equation}
{h(k_\perp)=\frac{2 k_\perp M_0}{{k_\perp}^2+M_0^{2}}} 
\end{equation}
where 
$M_0=\sqrt{\langle{k_\perp^2}\rangle}$ and $M_1$ are best fit parameters. Here, we will present the results for Model I 
only. The results for Model II and a comparison of the two models can be found in Ref.\cite{Godbole:2012bx}.  
 For ${\mathcal N}_g(x)$ also, we have used two kinds of parametrizations
\cite{Boer-PRD69(2004)094025}
\begin{itemize}
\item[(a)] ${\mathcal N}_g(x)=\left( {\mathcal N}_u(x)+
{\mathcal N}_d(x) \right)/2 \;$,
\item[(b)] ${\mathcal N}_g(x)={\mathcal N}_d(x) \;$,
\end{itemize}
where ${\mathcal N}_u(x)$ and ${\mathcal N}_d(x) \;$ are the normalizations for u and d quarks given by\cite{Boer-PRD69(2004)094025} 
\begin{equation} 
{\mathcal N}_f(x) = N_f x^{a_f} (1-x)^{b_f} \frac{(a_f + b_f)^{(a_f +
b_f)}}{{a_f}^{a_f} {b_f}^{b_f}} 
\label{siversx} 
\end{equation}
Here, $a_f, b_f$ and $N_f$ are best fit parameters fitted from new HERMES and COMPASS data\cite{Anselmino:2011gs} fitted at 
$\langle Q^2 \rangle = 2.4 GeV^2$ as given below: 
\begin{equation} 
N_u = 0.4, a_u=0.35, b_u =2.6  \nonumber\\
\end{equation}
\begin{equation}
N_d = -0.97,  a_d = 0.44, b_d=0.90 \nonumber\\
\end{equation}
\begin{equation}
M_1^2=0.19.\nonumber \\
\end{equation}\\
\\
\\
\\
We have estimated SSA using both Model I and II and parameterizations (a) and (b). The detailed results can be found in Ref.~\cite{Godbole:2012bx}. 
 
\section{TMD Evolution of PDF's and Sivers Function}

Early  phenomenological fits of Sivers function   were performed using experimental data at fixed scales and estimates of asymmetry were also 
performed either neglecting QCD 
evolution of TMD 	 PDF's or by applying DGLAP evolution only to the collinear part of TMD parametrization. In our earlier estimates 
of asymmetry in Ref.\cite{Godbole:2012bx} also, we have assumed the $Q^2$-dependence of PDF's and the Sivers function to be of the form, 
\begin{equation}
f_{g/p}(x,k_{\bot};Q)=f_{g/p}(x;Q)\frac{1}{\pi\langle k^{2}_{\bot}\rangle} 
e^{-k^{2}_{\bot}/\langle{k^{2}_{\bot}\rangle}} 
\label{gauss3}
\end{equation} 
and 
\begin{eqnarray}
\Delta^Nf_{g/p\uparrow}(x,k_{\perp};Q) = 2{\mathcal N}_g(x) f_{g/p}(x;Q) \nonumber \\
\times \sqrt{2e}\>\frac{k_{\bot}}{M_1}\>\frac{1}{\pi\langle k^{2}_{\perp}\rangle} 
e^{{-k^{2}_\perp}/\langle{k^{2}_{\perp}\rangle_S}} 
\label{dnf1}
\end{eqnarray}
where $\langle k^{2}_{\bot}\rangle=0.25~GeV^2$.  Note that the $Q^2$ dependence 
of PDF comes from collinear PDF $f_{g/p}(x;Q)$ only which have been evolved using DGLAP evolution. 
More recently, energy evolution of TMD's has been studied by various authors  and a TMD evolution 
formalism has been developed and implemented 
\cite{Collins:2011book, Aybat:2011zv}. 
TMD evolution is more complicated as compared to collinear counterpart 
because unlike collinear distributions TMDs have rapidity divergences in addition to collinear singularities. 
Thus TMD evolution describes how the form of distribution changes and also how the 
width changes in momentum space. 
 A strategy to extract Sivers function from SIDIS data taking into account the TMD $Q^2$ evolution 
has been proposed
\cite{Anselmino:2012aa}.
We have estimated SSA in electroproduction of $J/\psi$  taking into account this strategy.
In this formalism, the $Q^2$ dependence of PDF's is given by 
\begin{equation}
f_{q/p}(x,k_\perp;Q)=f_{q/p}(x,Q_0)\; R(Q,Q_0) \; 
\frac{ e^{-k_\perp^2/w^2}}{\pi\,w^2} \>,\label{unp-gauss-evol} 
\end{equation}
where, $f_{q/p}(x,Q_0)$ is the usual integrated PDF evaluated at the initial 
scale $Q_0$ 
and  $w^2 \equiv w^2(Q,Q_0)$ is the ``evolving'' 
Gaussian width, defined as
\begin{equation}
w^2(Q,Q_0)=\langle k_\perp^2\rangle + 2\,g_2 \ln \frac{Q}{Q_0}\> \cdot \label{wf}.
\end{equation} 
$R(Q,Q_0)$ is the limiting value of a function $R(Q, Q_0, b_T)$ that drives the $Q^2$-evolution of TMD's 
in coordinate space  
 and is driven by 
\begin{eqnarray}
 R(Q, Q_0, b_T)
\equiv 
\exp \{ \ln \frac{Q}{Q_0} \int_{Q_0}^{\mu_b} \frac{\rm d \mu'}{\mu'} \gamma_K(\mu') \nonumber \\
+\int_{Q_0}^Q \frac{\rm d \mu}{\mu} 
\gamma_F \left( \mu, \frac{Q^2}{\mu^2} \right)\} 
\> \cdot  
\end{eqnarray}
where $b_T$ is the parton impact parameter, \\
\begin{equation}
b_*(b_T) \equiv \frac{b_T}{\sqrt{1 + b_T^2/b_{\rm max}^2}} , 
\quad\quad\quad
\mu_b = \frac{C_1}{b_*(b_T)} \> 
\end{equation}

with $C_1=2 e^{-\gamma_E}$ where $\gamma_E=0.577$, 
$b_*\rightarrow b_{max}$. 

 $\gamma_F$ and $\gamma_K$ are anomalous dimensions which are given at $O(\alpha_s)$ by
\begin{equation}
\gamma_F(\mu ; \frac{Q^2}{\mu^2}) = \alpha_s(\mu) \, \frac{C_F}{\pi}
\left(\frac{3}{2} - \ln \frac{Q^2}{\mu^2} \right) 
\label{gammaF} 
\end{equation}

\begin{equation}
\gamma_K(\mu) = \alpha_s(\mu) \, \frac{2 \, C_F}{\pi} \> \cdot
\label{gammaK} 
\end{equation}
 \ In the limit $b_T \rightarrow \infty$ , $R(Q, Q_0, b_T )\rightarrow R(Q, Q_0)$. 

\section{Numerical Estimates}

We have estimated SSA in electroproduction of $J/\psi$ for JLab, HERMES, COMPASS and eRHIC 
energies. Our earlier calculation of asymmetry\cite{Godbole:2012bx} had taken into account energy evolution of 
PDF's and Sivers function using DGLAP evolution.  The details can be found in Ref.\cite{Godbole:2012bx}. 
 
 In Figs.1-5, we have presented a comparison of SSA's calculated using DGLAP  evolution and TMD 
 evolution of TMD PDF's at various energies for Model I with parametrization (a). For TMD evolved Sivers function, we have used the parameter set fitted at $Q_0=1$~GeV given in Ref. \cite{Anselmino:2012aa}
\begin{eqnarray}
N_u = 0.75, \ N_d = -1.00,  \nonumber \\ 
b=4.0,\, a_u=0.82, \, a_d = 1.36,   \nonumber \\
{M_1}^2=0.34~GeV^2, g_2 = 0.68 \;.\label{tmd-parm} 
\end{eqnarray}
It is found that the asymmetry is substantially reduced 
 in all cases when TMD evolution of PDF's and Sivers function is taken into account.  Here, we have used parametrization (a) for our estimates. 
 A more detailed analysis with parametrization (b) and a comparison of the two parametrizations as well as of various parameter sets can be 
 found in Ref\cite{Godbole:2013bca}.


\section{Summary}
Transverse SSA in electroprduction of $J/\psi$ has been calculated  using color evaporation model of charmonium production.  
A TMD factorization formalism has been used first with DGLAP evolved PDF's  and then with TMD evolved PDF's and Sivers function. 
Sizable asymmetry is predicted at  energies of JLab, HERMES, COMPASS and eRHIC experiments in both cases. However, it is found that there is a 
substantial reduction in asymmetry when TMD evolution is taken into account. 
Substantial magnitude of asymmetry indicate that it may be worthwhile  to look at SSA's in charmonium production both from the point of view of comparing
different models of charmonium production as well as comparing the different models of gluon 
Sivers function. It is also clear that TMD evolution effects are substantial and one must take them into account for accurate predictions. \\

{\bf ACKNOWLEDGEMENTS}\\
I would like to thank the organizers of LC2012 Delhi for their kind hospitality. I would also like to thank DAE BRNS, India 
 for financial support during this project under the grant No. 2010/37P/47/BRNS.   

\begin{figure}[h]
\includegraphics[scale=0.52]{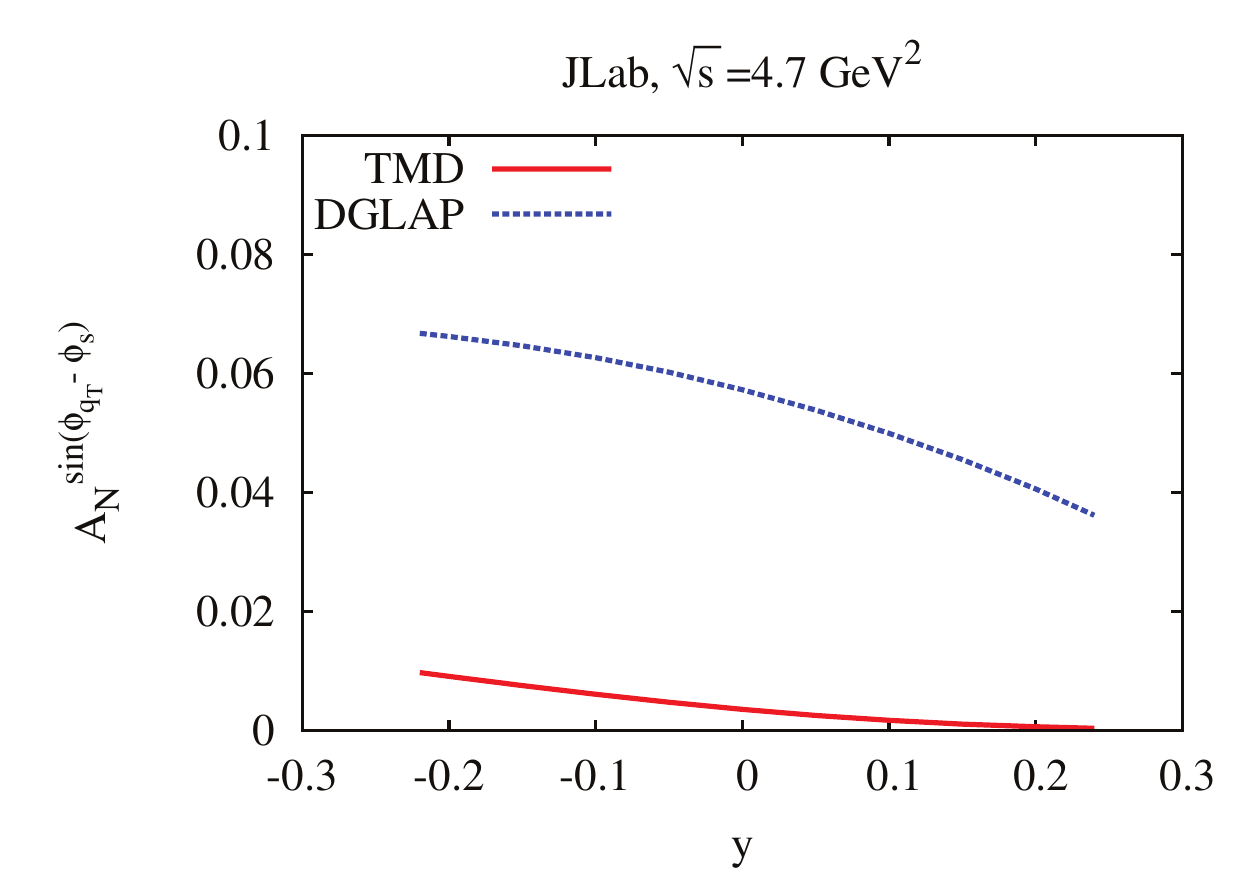} 
\includegraphics[scale=0.52]{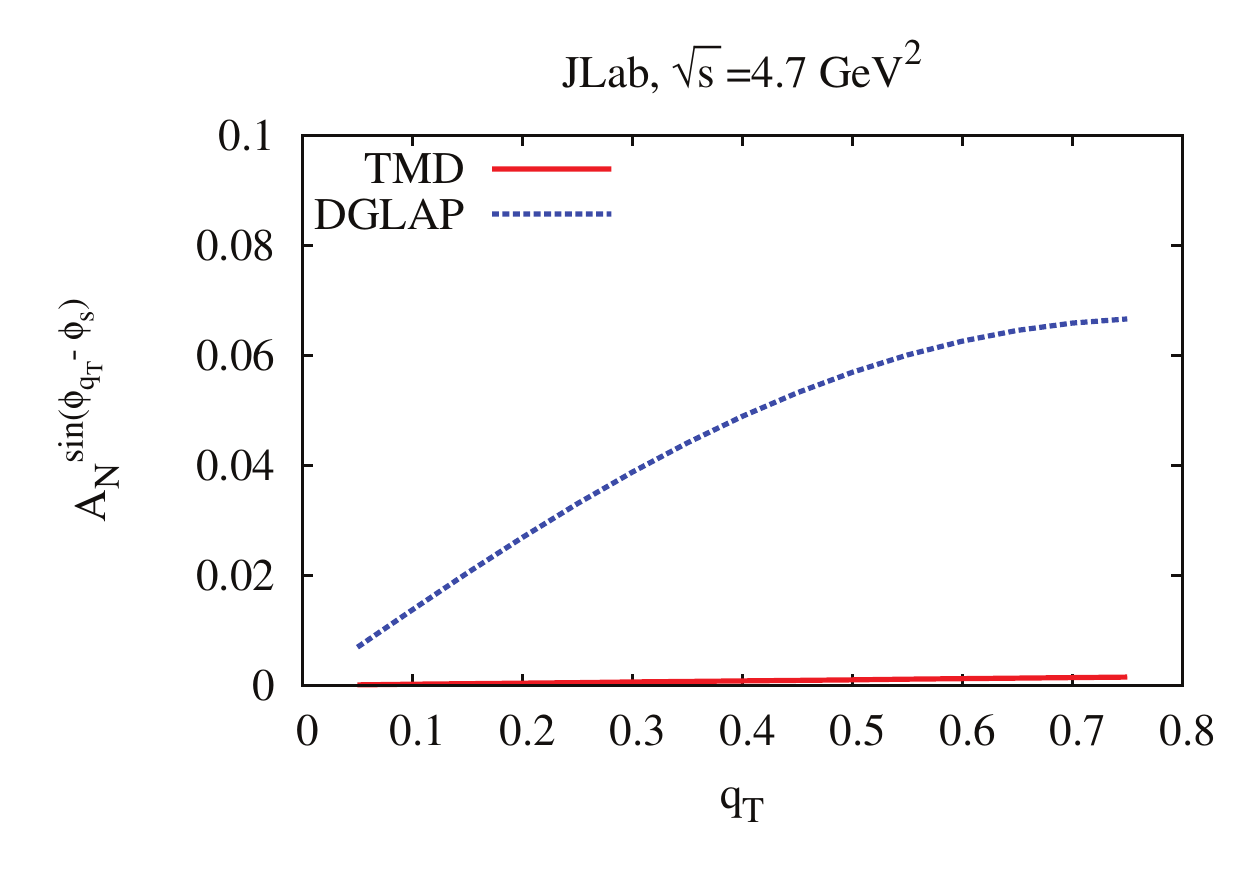}
\caption{The Sivers asymmetry $A_N^{\sin({\phi}_{q_T}-\phi_S)}$ for $e+p^\uparrow \to  e+J/\psi +X $
at JLab energy ($\sqrt{s} = 4.7$~GeV) as a function of y (top panel) and $q_T$ (bottom panel)
for parametrization (a).}
  \end{figure}
  \begin{figure}[h]
\includegraphics[scale=0.52]{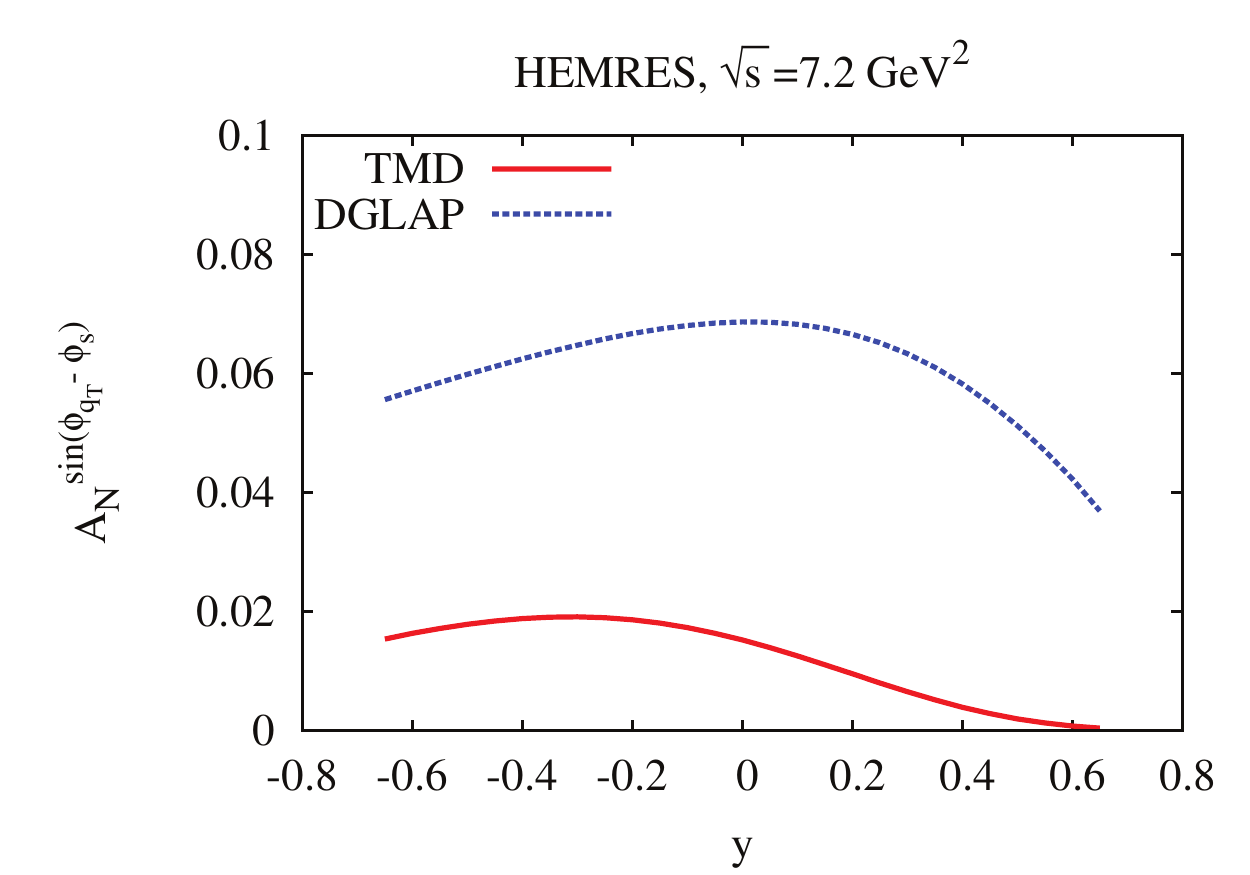} 
\includegraphics[scale=0.52]{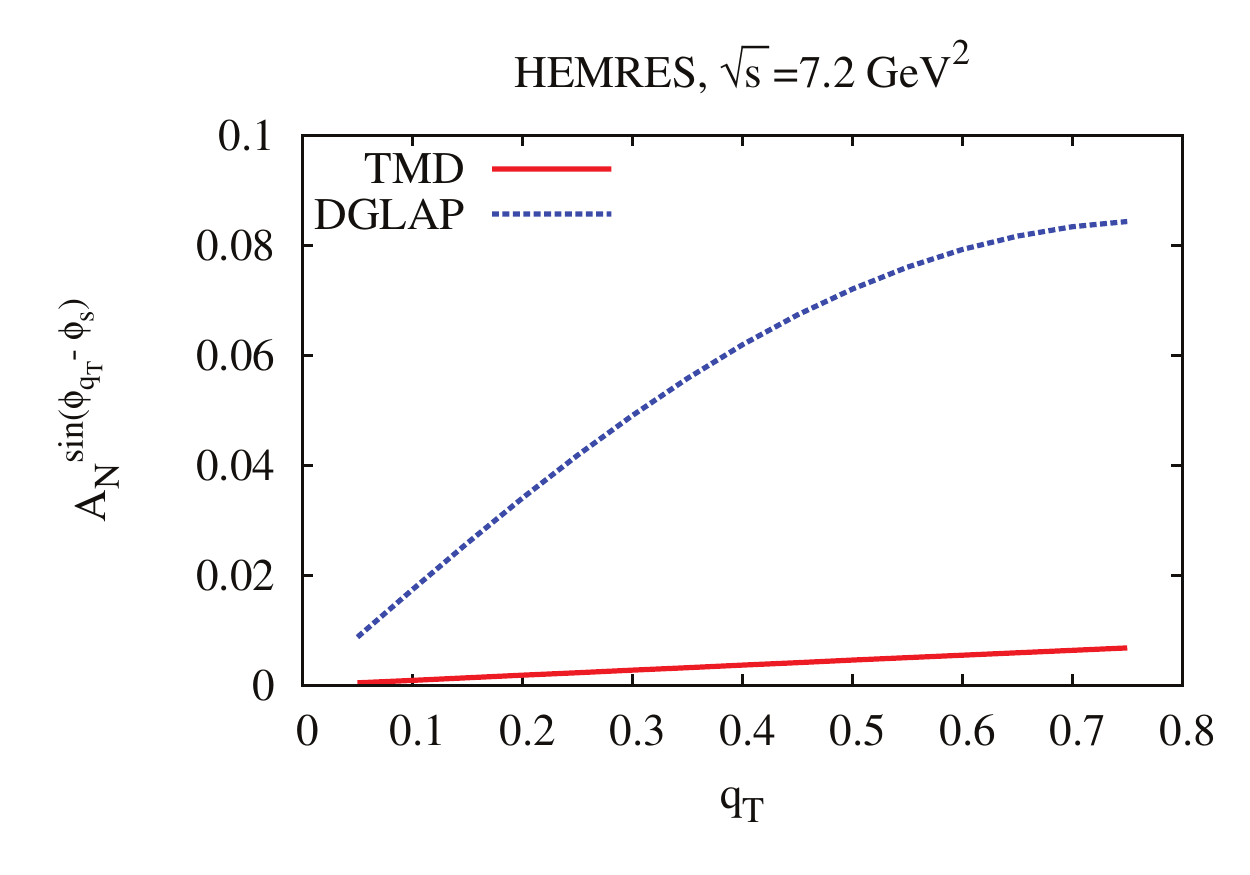}
\caption{The Sivers asymmetry $A_N^{\sin({\phi}_{q_T}-\phi_S)}$ for $e+p^\uparrow \to  e+J/\psi +X $
at HEMRES energy ($\sqrt s = 7.2$~GeV) as a function of y (top panel) and $q_T$ (bottom panel)
for parametrization (a).}
\end{figure}
\begin{figure}[h]
\includegraphics[scale=0.52]{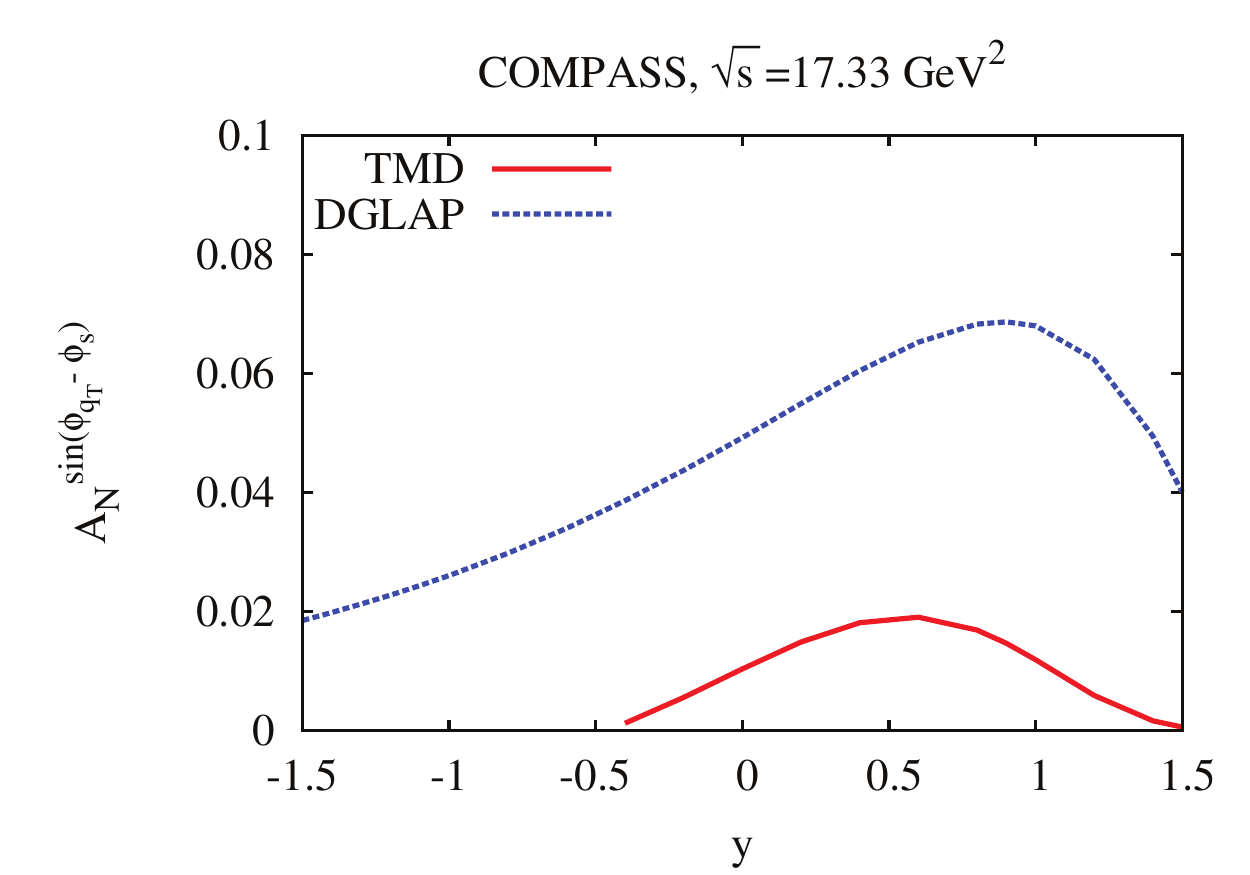} 
\includegraphics[scale=0.52]{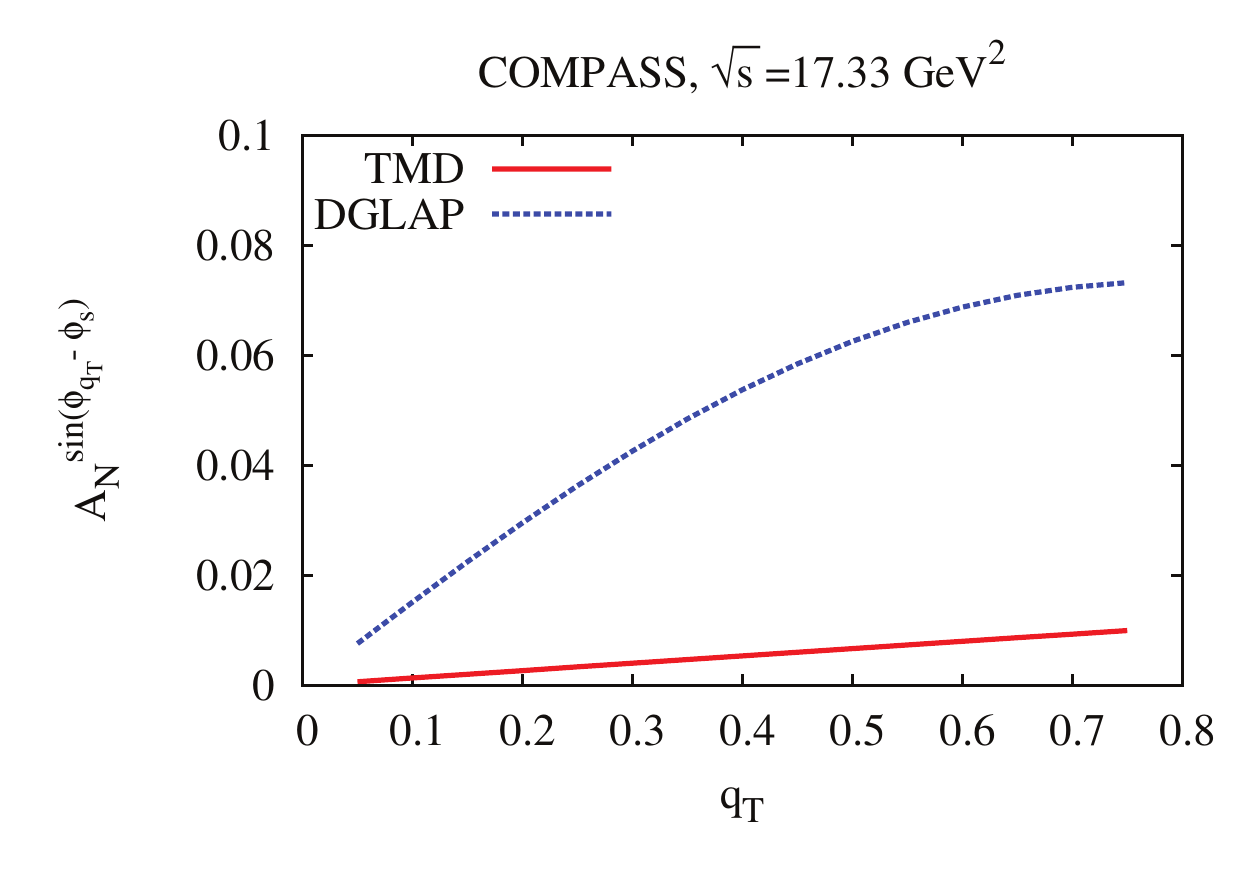}
\caption{The Sivers asymmetry $A_N^{\sin({\phi}_{q_T}-\phi_S)}$ for $e+p^\uparrow \to  e+J/\psi +X $
at COMPASS energy ($\sqrt s = 17.33$~GeV) as a function of y (top panel) and $q_T$ (bottom panel)
for parametrization (a).}
 \end{figure}
 \begin{figure}[h]
\includegraphics[scale=0.52]{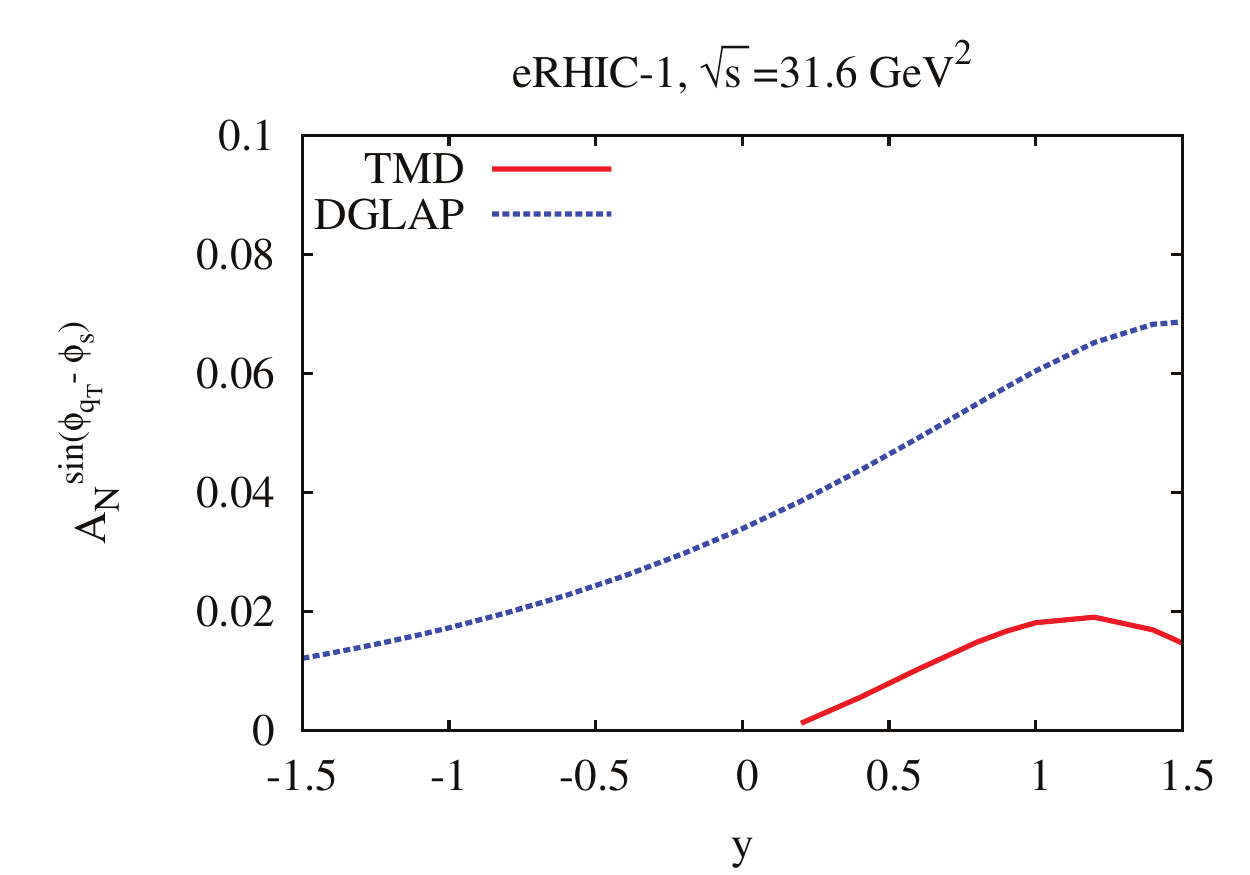} 
\includegraphics[scale=0.52]{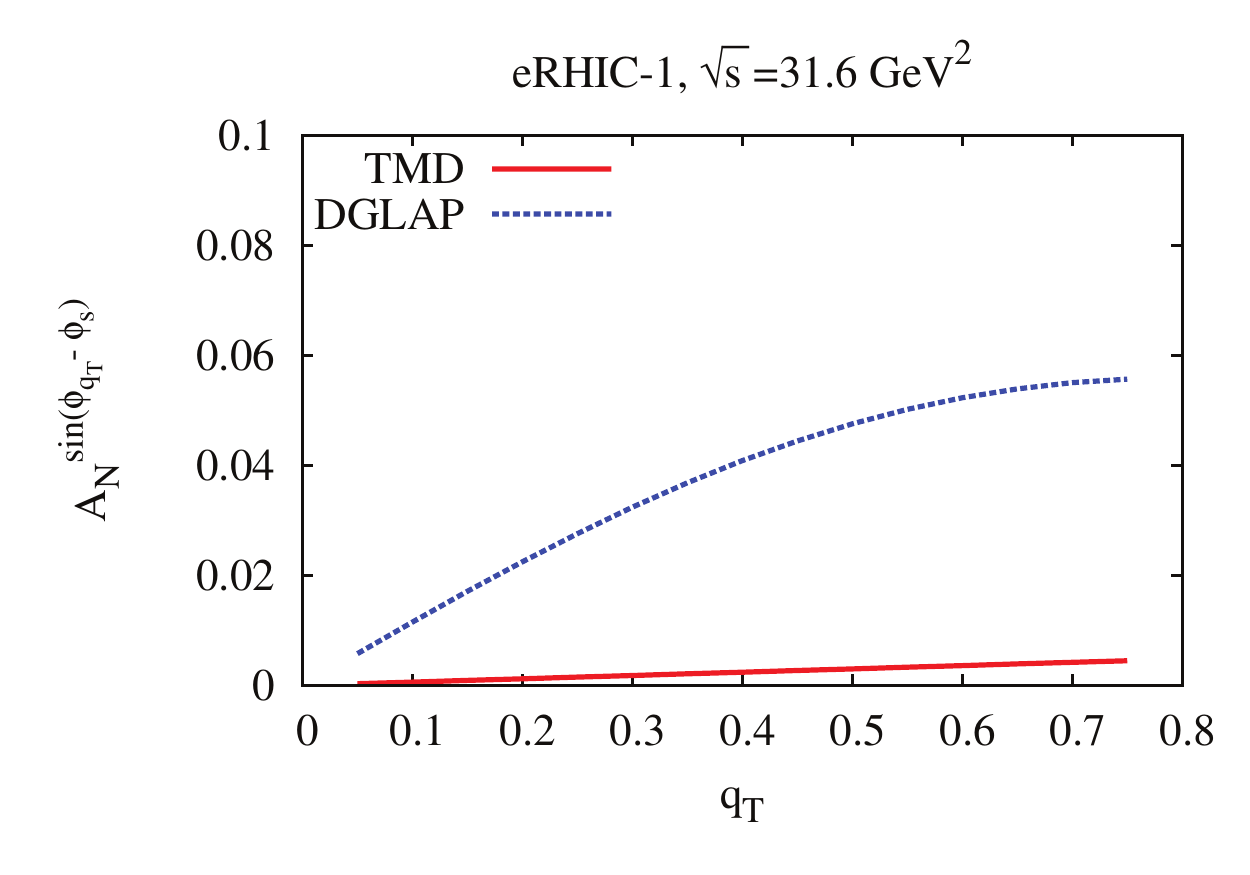}
\caption{The Sivers asymmetry $A_N^{\sin({\phi}_{q_T}-\phi_S)}$ for $e+p^\uparrow \to  e+J/\psi +X $
at eRHIC-1 energy ($\sqrt{s}=31.6$~GeV) as a function of y (top panel) and $q_T$ (bottom panel)
for parametrization (a).}
\end{figure}
\begin{figure}[h]
\includegraphics[scale=0.52]{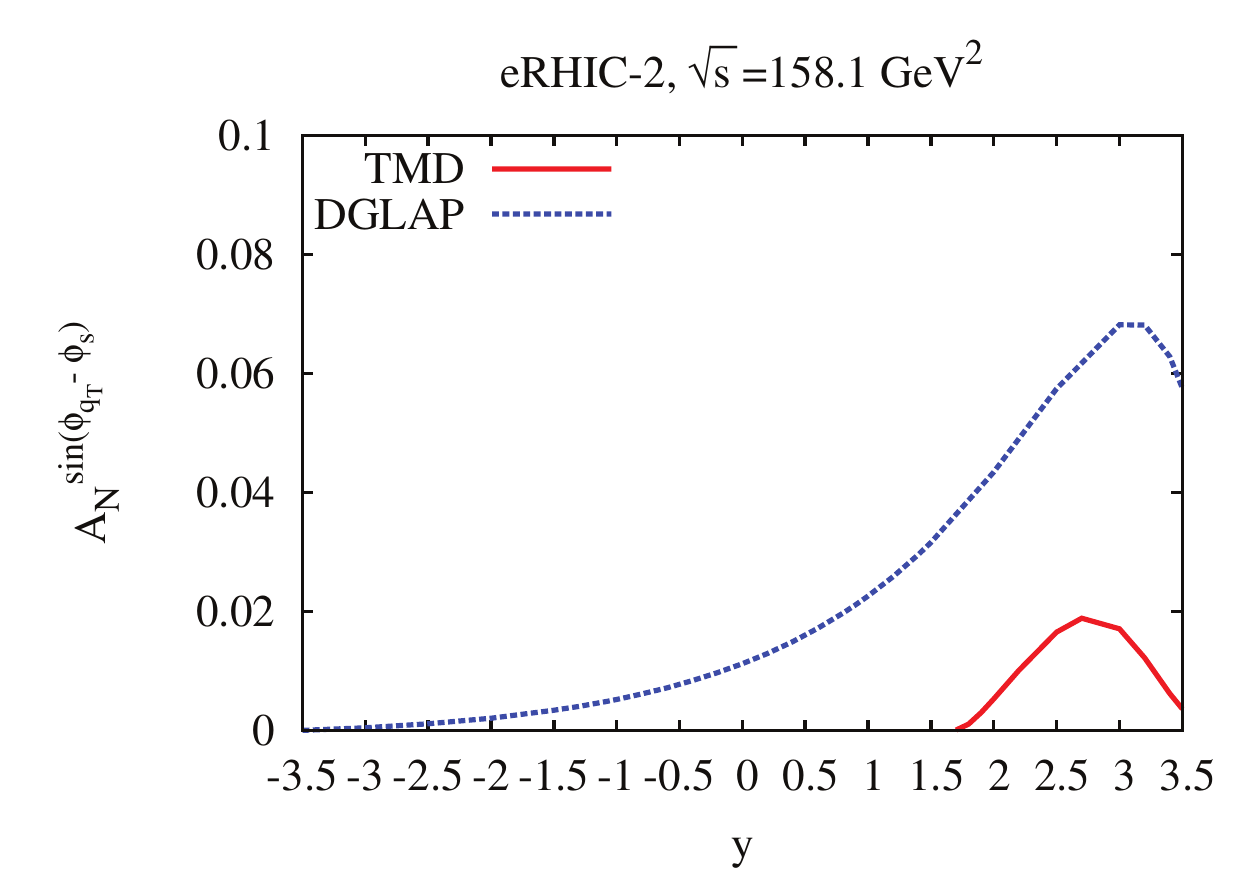} 
\includegraphics[scale=0.52]{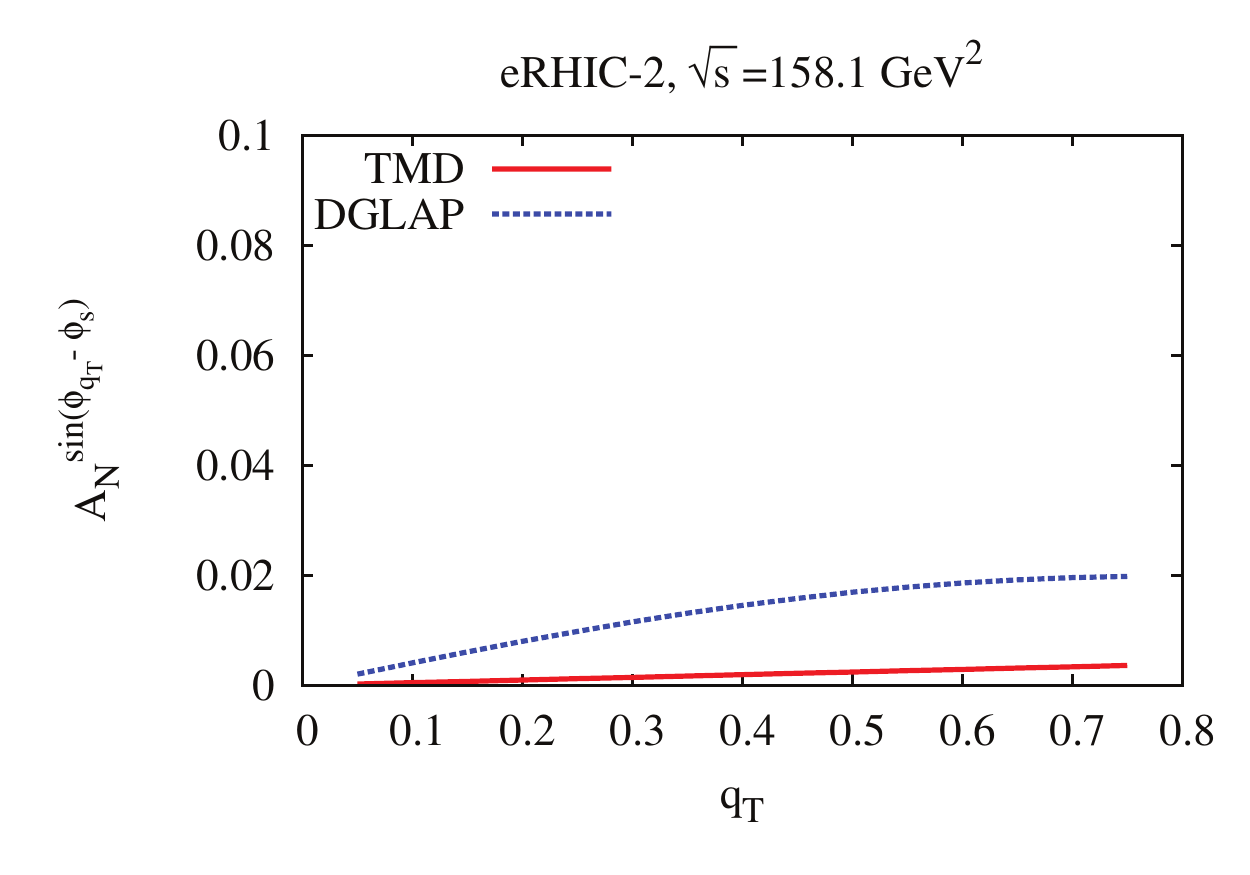}
\caption{The Sivers asymmetry $A_N^{\sin({\phi}_{q_T}-\phi_S)}$ for $e+p^\uparrow \to  e+J/\psi +X $
at eRHIC-2 energy ($\sqrt{s}=158.1$~GeV) as a function of y (top panel) and $q_T$ (bottom panel)
for parametrization (b).}
\end{figure}


\small
\begin{thebibliography}{abcdef99}
\bibitem{AdamsBravar1991}   D.~L.~Adams {\it et al.}  [FNAL-E704 Collaboration],
  Phys.\ Lett.\ B {\bf 264}, 462 (1991);
  A.~Bravar {\it et al.}  [Fermilab E704 Collaboration],
  Phys.\ Rev.\ Lett.\  {\bf 77}, 2626 (1996).

\bibitem{KruegerAllogower1999} K.~Krueger, C.~Allgower, T.~Kasprzyk, H.~Spinka, D.~Underwood, A.~Yokosawa, G.~Bunce and H.~Huang {\it et al.},
  Phys.\ Lett.\ B {\bf 459}, 412 (1999);
  C.~E.~Allgower, K.~W.~Krueger, T.~E.~Kasprzyk, H.~M.~Spinka, D.~G.~Underwood, A.~Yokosawa, G.~Bunce and H.~Huang {\it et al.},
  Phys.\ Rev.\ D {\bf 65}, 092008 (2002).

\bibitem{Hermes} A.~Airapetian {\it et al.}  [HERMES Collaboration],
  Phys.\ Rev.\ Lett.\  {\bf 84}, 4047 (2000)
  [hep-ex/9910062];
  Phys.\ Rev.\ D {\bf 64}, 097101 (2001)
  [hep-ex/0104005].

\bibitem{Compass}   V.~Y.~Alexakhin {\it et al.}  [COMPASS Collaboration],
  Phys.\ Rev.\ Lett.\  {\bf 94}, 202002 (2005)
  [hep-ex/0503002].

\bibitem{alesio-review} 
  U.~D'Alesio and F.~Murgia,
  Prog.\ Part.\ Nucl.\ Phys.\  {\bf 61}, 394 (2008)
  [arXiv:0712.4328 [hep-ph]].

\bibitem{Sivers1990} D.~W.~Sivers,
  Phys.\ Rev.\ D {\bf 41}, 83 (1990);
  Phys.\ Rev.\ D {\bf 43}, 261 (1991).

\bibitem{Anselmino:2008sga} 
  M.~Anselmino, M.~Boglione, U.~D'Alesio, A.~Kotzinian, S.~Melis, F.~Murgia, A.~Prokudin and C.~Turk,
  Eur.\ Phys.\ J.\ A {\bf 39}, 89 (2009).
  [arXiv:0805.2677 [hep-ph]].

\bibitem{Boer-PRD69(2004)094025}D.~Boer and W.~Vogelsang,
  Phys.\ Rev.\ D {\bf 69}, 094025 (2004)
  [hep-ph/0312320].
 
     
  \bibitem{Anselmino2004}M. Anselmino, M. Boglione, U. D'Alesio, E.  Leader and F. Murgia, 
\PR {D 70}, 074025(2004); hep-ph/0407100.


 \bibitem{feng} F.~Yuan,
  Phys.\ Rev.\ D {\bf 78}, 014024 (2008)
  [arXiv:0801.4357 [hep-ph]].

  
  \bibitem{Brodsky2002}S.~J.~Brodsky, D.~s.~Hwang and I.~Schmidt, Phys.Lett. B530,99(2002);Nucl.Phys.B642,344(2002).
  
  \bibitem{sing} E. L. Berger, D. Jones, Phys. Rev {\bf D 23 }, 1521 (1981); R. Bair, R. R\"uckl, Phys. Lett. 
{\bf B 102}, 364 (1981), Nucl. Phys. {\bf B 201}, 1 (1982).


\bibitem{hal} F. Halzen, Phys. Lett. {\bf B 69}, 105 (1997); F. Halzen, S. Matsuda, Phys. Rev. 
{\bf D 17}, 1344 (1978), H. Fritsch, Phys. Lett. {\bf B 67}, 217 (1977),  O. J. P. Eboli, E. M. Gregores, F. Halzen, Phys. Rev. {\bf D
67}, 054002 (2003). 

\bibitem{cem0}M.~B.~Gay Ducati and C.~Brenner Mariotto,
  Phys.\ Lett.\ B {\bf 464}, 286 (1999)
  [hep-ph/9908407].


\bibitem{octet} G. T. Bodwin, E. Braaten, G. P. Leapage, Phys. Rev.
 {\bf D 43}, 1914 (1992). 
 
 
\bibitem{Godbole:2012bx} 
  R.~M.~Godbole, A.~Misra, A.~Mukherjee and V.~S.~Rawoot,
  Phys.\ Rev.\ D {\bf 85}, 094013 (2012)
  [arXiv:1201.1066 [hep-ph]].

 
 \bibitem{cem1} O.~J.~P.~Eboli, E.~M.~Gregores and F.~Halzen,
  hep-ph/0211161.
 
 \bibitem{cem2}
  G.~T.~Bodwin, E.~Braaten and J.~Lee,
  Phys.\ Rev.\ D {\bf 72}, 014004 (2005)
  [hep-ph/0504014].

\bibitem{wwf1}C. F. Weizsacker, \ZP {88} (1934),  E. J. Williams, \PR {45} 729(1934).
  

\bibitem{ww} B.~A.~Kniehl,
  Phys.\ Lett.\ B {\bf 254}, 267 (1991).

 \bibitem{gr78}M.~Gluck and E.~Reya,
  Phys.\ Lett.\ B {\bf 79}, 453 (1978).

\bibitem{vogelsang-weight} W. Vogelsang, and F. Yuan, \PR {\bf D 72}, 054028
 (2005); arXiv:hep-ph/0507266, J. C. Collins et al., Phys. Rev. D 73, 094023 (2006).
 
\bibitem{Anselmino2009} M.~Anselmino, M.~Boglione, U.~D'Alesio, S.~Melis, F.~Murgia and A.~Prokudin,
  Phys.\ Rev.\ D {\bf 79}, 054010 (2009)
  [arXiv:0901.3078 [hep-ph]].
  
\bibitem{Anselmino:2011gs} 
  M.~Anselmino, M.~Boglione, U.~D'Alesio, S.~Melis, F.~Murgia and A.~Prokudin,
[hep-ph/1107.4446]

  \bibitem{Collins:2011book}
  J.~C.~Collins, Foundations of Perturbative QCD, Cambridge Monographs 
  on Particle Physics, Nuclear Physics and Cosmology, No. 32, Cambridge 
  University Press, Cambridge, 2011.

\bibitem{Aybat:2011zv} 
  S.~M.~Aybat and T.~C.~Rogers,
  Phys.\ Rev.\ D {\bf 83}, 114042 (2011)
  [arXiv:1101.5057 [hep-ph]],
  S.~M.~Aybat, J.~C.~Collins, J.~-W.~Qiu and T.~C.~Rogers,
  arXiv:1110.6428 [hep-ph],
  S.~M.~Aybat, A.~Prokudin and T.~C.~Rogers,
  arXiv:1112.4423 [hep-ph].

\bibitem{Anselmino:2012aa} 
  M.~Anselmino, M.~Boglione and S.~Melis,
  Phys.\ Rev.\ D {\bf 86}, 014028 (2012)
  [arXiv:1204.1239 [hep-ph]].
  
\bibitem{Godbole:2013bca} 
  R.~M.~Godbole, A.~Misra, A.~Mukherjee and V.~S.~Rawoot,
  Phys.\ Rev.\ D {\bf 88}, 014029 (2013)
  [arXiv:1304.2584 [hep-ph]].
\end{thebibliography}
\end{document}